\documentclass{ws-p8-50x6-00}

\font\eightrm=cmr10 scaled 800
\def\pbar{\overline p}
\def\taup{\tau_p}
\def\tpbar{\tau_{\pbar}}
\def\lesssim{\mathrel{\vcenter{\offinterlineskip\halign{\hfil
$\displaystyle##$\hfill\cr<\cr\sim\cr}}}}
\def\gtrsim{\mathrel{\vcenter{\offinterlineskip\halign{\hfil
$\displaystyle##$\hfill\cr>\cr\sim\cr}}}}

\def\ApJ{\em Astrophys. J.}
\def\PR{\em Phys. Rep.}
\def\NPA{\em Nucl. Phys. A}

\begin{document}

\title{Cosmic Ray Antiprotons}

\author{Dallas C. Kennedy}

\address{Department of Physics, University of Florida, Gainesville FL 32611,
USA\\E-mail: kennedy@phys.ufl.edu $\bf{\ \bullet\bullet}$ WWW: http://www.phys.ufl.edu/\~{}kennedy}

%%%%%%%%%%%%%%%%%%%%%%%%%%%%%%%%%%%%%%%%%%%%%%%%%%%%%%%%%%%%%%
% You may repeat \author \address as often as necessary      %
%%%%%%%%%%%%%%%%%%%%%%%%%%%%%%%%%%%%%%%%%%%%%%%%%%%%%%%%%%%%%%

\maketitle

\abstracts{
Cosmic ray antiprotons have been detected
for over 20 years and are now measured reliably.  Standard particle and
astrophysics predict a conventional spectrum and abundance of secondary
antiprotons consistent with all current measurements.  These
measurements place limits on exotic Galactic antiproton sources and
non-standard antiproton properties.  Complications
arise, particularly at low energies, with heliospheric modulation of cosmic
ray fluxes and production of standard secondaries from $A >$ 1 nuclear
targets.  Future experiments and theoretical developments are discussed.}

\section{General Properties of Cosmic Rays}

\begin{figure}
%\figurebox{10pc}{10pc}{Hep00-3Fig1.eps} % to have a box alone
\epsfxsize=25pc % will enlarge or reduce the postscript figures based on the xsize
\epsfbox{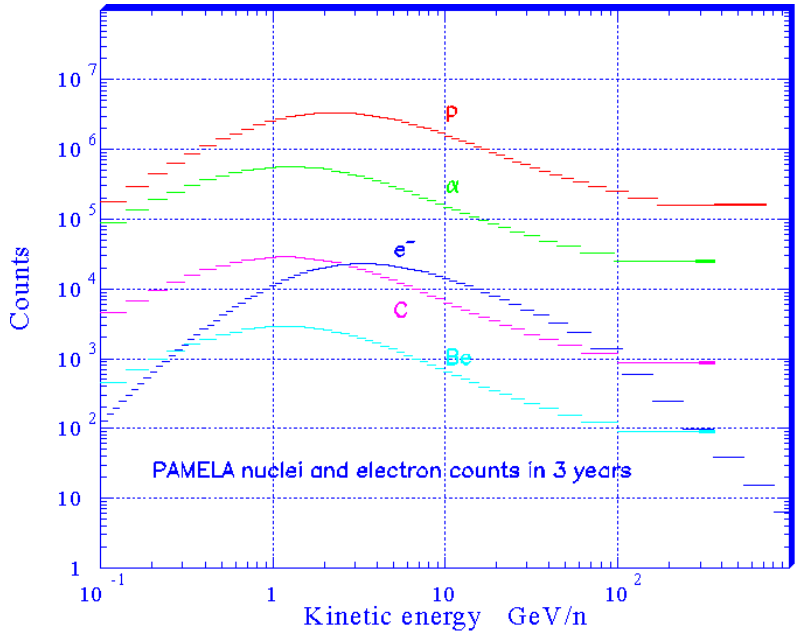} % postscript image file name
\caption{Typical cosmic ray matter spectrum: isotopic count spectra expected for
orbital PAMELA instrument (see text) over 3 years = 95 Msec.  Isotopes
shown: H, He, Be, and C, as well as electrons 
$e^-$.~\protect\cite{pHeSpectra}\label{fig:pSpec}}
\end{figure}

Cosmic rays are high-energy $p$, $\pbar$, nuclei, and $e^\pm$
in interplanetary and interstellar (IS) space.  The dominant
component consists of protons (hydrogen H) with a smaller admixture of
heavier nuclei, especially He (Figure~\ref{fig:pSpec}).
Antiprotons $(\pbar)$ occur at an
abundance of $10^{-4,-5}$ times that of $p$.  These energetic particles,
with kinetic energy $K >$ 10 MeV, are Galactic in origin, not to be confused
with the much denser solar wind plasma, with much lower $K$, streaming from
the Sun.\footnote{Also ignored here are ``pickup ions'' or anomalous Galactic cosmic rays, 
neutral IS atoms which drift into the solar system and are then ionized by solar UV
radiation.}

The relative element abundances in cosmic rays (CRs) indicate they originate
in the IS medium, where they are ionized and accelerated, probably by
supernova shocks.  (Recent measurements all but rule out an
origin in supernova ejecta proper.)  Such accelerated, pre-existing
nuclei are CR {\it primaries}.~\cite{ACE}  Once accelerated to high energies, the
primaries induce the production of further CRs, the {\it secondaries},
in the IS medium and at local sites in the Galaxy.  (The terms ``primaries''
and ``secondaries'' are also used in a completely different sense: CR
primaries are the CRs that strike the top of the Earth's atmosphere, the
secondaries the induced CR shower propagating into the lower atmosphere.)
Secondaries include $e^+$, $\pbar$, and certain nuclear isotopes.
As some of these isotopes are unstable, their populations must be continually
replenished to maintain their observed abundances.

\subsection{Cosmic Ray Antiprotons}

Standard secondary CR antiprotons are produced by the process $pA\rightarrow\pbar X$,
with $p$ = high-energy CR, $A$ = IS medium nucleus of atomic weight $A$, and
$X$ = anything consistent with charge and baryon number $(B)$ conservation.
The threshold channel is $pA\rightarrow p\pbar pA$, with threshold $E_p$ =
$(3+4/A)m_p$.  (The nucleus $A$ can break up without significantly changing
the dynamics.)  The dominant case is H, $A$ = 1, with threshold $E_p$ =
$7m_p$.  The only other significant contribution comes from He target nuclei.
By number, the IS medium is $\simeq$ 93\% H and 7\% 
He.~\cite{GaisLevy,Stephens,WebbPot,GaisSch,Gaisser}

The secondary $\pbar$'s subsequently propagate in the Galaxy and are subject
to a variety of elastic (scattering, including energy-loss) and inelastic
(annihilation and extra-Galactic leakage) processes.  Leakage is the dominant 
loss; the Galactic storage time $\sim$ 13 Myr as inferred from
the abundance of unstable CR isotopes.  Energy loss shifts the $\pbar$ spectrum
without changing their number.  Some uncertainty is unavoidable in models of
Galactic propagation, including H and He abundances, as well as the Galaxy's 
highly tangled, stochastic magnetic field
$B_{\rm Gal}\sim$ 0.3 nT and small wind $V_{\rm Gal}\lesssim$ 20 km sec$^{-1}$ 
(a superposition of many stellar and supernova winds) .  The field and wind 
control the diffusion of CRs into intergalactic space and are fairly well
constrained by measurements of unstable CR isotopes.
But more complicated transport mechanisms are possible, including 
reacceleration shocks and variation of the Galactic 
geometry.~\cite{Gaisser,WebberLeeGupta,Longair,Chard}

\begin{figure}
%\figurebox{20pc}{15pc}{} % to have a box alone
\epsfxsize=20pc % will enlarge or reduce the postscript figures based on the xsize
\epsfbox{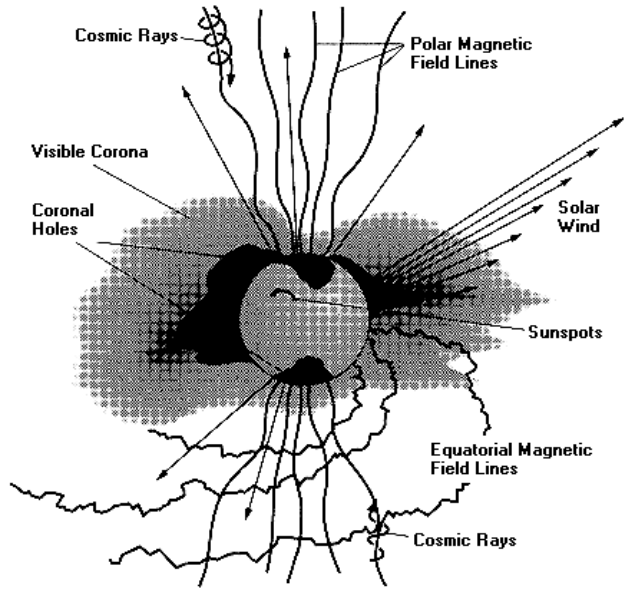} % postscript image file name
\caption{Galactic cosmic rays propagate into heliosphere by spiraling along heliomagnetic
field lines, but the field itself also fluctuates randomly.  Solar wind also convects and
decelerates incoming cosmic rays.~\protect\cite{Ulysses}\label{fig:helioA}}
\end{figure}

The CR fluxes measured at the top of the Earth's atmosphere are modulated by their transport 
through the heliosphere, the Sun's magnetic sphere of influence (Figure~\ref{fig:helioA}).
The heliosphere consists of 
a solar wind $(V_W\simeq$ 400 km sec$^{-1}$ along the ecliptic plane, 700-800 km sec$^{-1}$ along the
solar axes) of $e^-$ and nuclei (mainly $p)$ carrying the embedded magnetic field $B_\odot$ 
(Figure~\ref{fig:helioA}).  
Near the Sun, the field falls off with heliocentric distance $r$ as $r^{-2}$, 
arising from frozen field flux transported radially outwards.  Since the Sun rotates, however,
the field is twisted into an Archimedean or {\em Parker spiral}, and
the field is predominantly azimuthal in the outer solar system, falling off more softly as
$r^{-1}$.  At one AU (AU = 149.5 Mkm, the Earth's orbit), the heliomagnetic field strength 
$B_\odot\simeq$ 5 nT.~\cite{Parker}

The CRs gyrate around the local ${\bf B}$ field lines.  
The solar field is not fully deterministic,
however: it is modified by episodic (practically random) shocks that cause
the CRs to diffuse along and across field lines, especially at times of solar magnetic
maximum (currently 2000-01 and periodically about every 11 years, when the heliomagnetic
field changes sign).  In addition, the wind both
imposes a macroscopic convective drift and performs work on the CRs (adiabatic 
deceleration), lowering their energies as they fight ``upstream''
into the inner solar system.  A realistic prediction of Earth-measured CR
fluxes must include these mechanisms, which are particularly important at lower
$K$ and affect oppositely-charged CRs differently.~\cite{transportOld,transportNew,Longair}  
Heliospheric {\it in situ}
measurements have been ongoing for four decades and have recently become
much better with the {\it IMP} and {\it Ulysses} space probes.~\cite{IMP,Ulysses}

\subsection{Exotic Sources of Cosmic Ray Antiprotons}

The density of IS matter $n_{\rm H}\sim$ 1 H atom cm$^{-3}$ and the 
known spectrum and abundance of CR $p$ primaries fix the predicted spectrum and abundance 
of $\pbar$ secondaries, if the $pA\rightarrow\pbar X$ cross section 
$\sigma (\pbar )$ is known.~\cite{Stephens}  Let $Q_{\pbar}(K)$ be the differential production 
rate (antiprotons cm$^{-3}$ MeV$^{-1}$ sec$^{-1}$); schematically,
\be
Q_{\pbar}(K) = \int\ dK^\prime\ n_p (K^\prime)v(p)n_{\rm H}\cdot
d\sigma (pp\rightarrow\pbar; K,K^\prime )/dK^\prime\quad .
\ee
The $pp$ process has been measured in laboratory experiments, and the $p+$He case can be 
inferred from the $pp$ cross section (but see subsection~\ref{sec:FutureTheo}).  The 
differential $\pbar$ abundance $n_{\pbar}(K)$ (antiprotons
cm$^{-3}$ MeV$^{-1}$) is related to $Q_{\pbar}(K)$ by $n_{\pbar}(K)$ = 
$\tau_{\rm eff}(K)\cdot Q_{\pbar}(K)$, where the effective Galactic residence time
\be
\frac{1}{\tau_{\rm eff}(K)} = \frac{1}{\tau_{\rm leak}(K)} + \frac{1}{\tau_{\rm ann}(K)}
+\cdots \quad,
\ee
summing over all loss mechanisms.  The sum is dominated by the first term, the
extra-Galactic diffusion rate.  The measured CR $\pbar$ flux is then related to $n_{\pbar}(K)$ 
through the transformation by heliospheric transport.  Variation of Galactic transport 
mechanisms modifies $\tau_{\rm eff}(K)$.  This picture is the basis for
the simple {\em Leaky Box Model}.  A more complex picture, with explicit spatial
dependence on Galactic geometry (inhomogeneous leaky disk model = ILDM), 
is possible and indeed necessary, because of measurements
of Galactic plane cosmic ray synchrotron radiation mapping the IS CR
distribution.~\cite{CRgammas,WebberLeeGupta,Chard}

\begin{figure}
%\figurebox{20pc}{15pc}{} % to have a box alone
\epsfxsize=20pc % will enlarge or reduce the postscript figures based on the xsize
\epsfbox{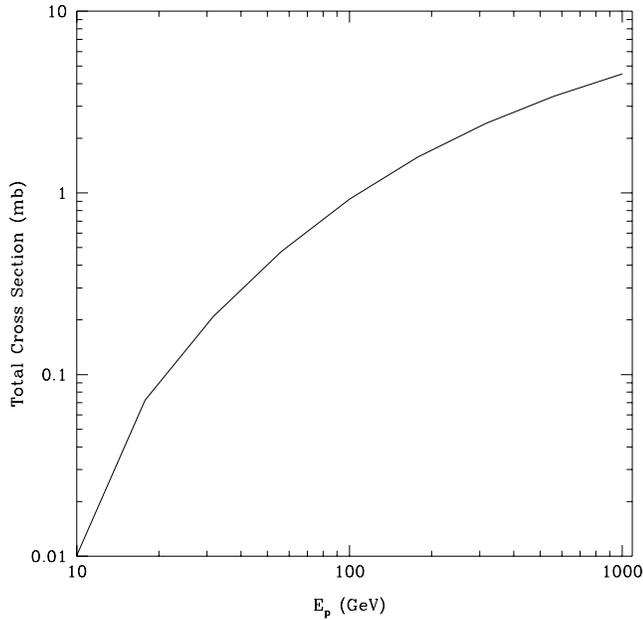} % postscript image file name
\caption{The total $pp\rightarrow\pbar X$ cross section as a function of incident $p$
total energy in target $p$ rest frame.  Based on measured cross section and
semianalytic representation by Stephens.~\protect\cite{Stephens}\label{fig:pbarProd}}
\end{figure}

The cross section $\sigma (\pbar )$ has a crucial property arising from its threshold 
at $E_p$ = 7$m_p$ (Figure~\ref{fig:pbarProd}).  The spectrum of outgoing $\pbar$'s
rises sharply from $K_{\pbar}$ = 0.  Since $n_p(K)$ falls off
rapidly (as $K^{-2.75}_p)$, $n_{\pbar}(K)$ falls off similarly at high $K_{\pbar}$, leaving a 
$\pbar$ secondary spectrum with a sharp rise to
a peak at $K\sim$ 2 GeV and falling off above that.  The lower threshold for
He targets enhances the low-$K$ spectrum somewhat.

\begin{figure}
%\figurebox{20pc}{15pc}{} % to have a box alone
\epsfxsize=25pc % will enlarge or reduce the postscript figures based on the xsize
\epsfbox{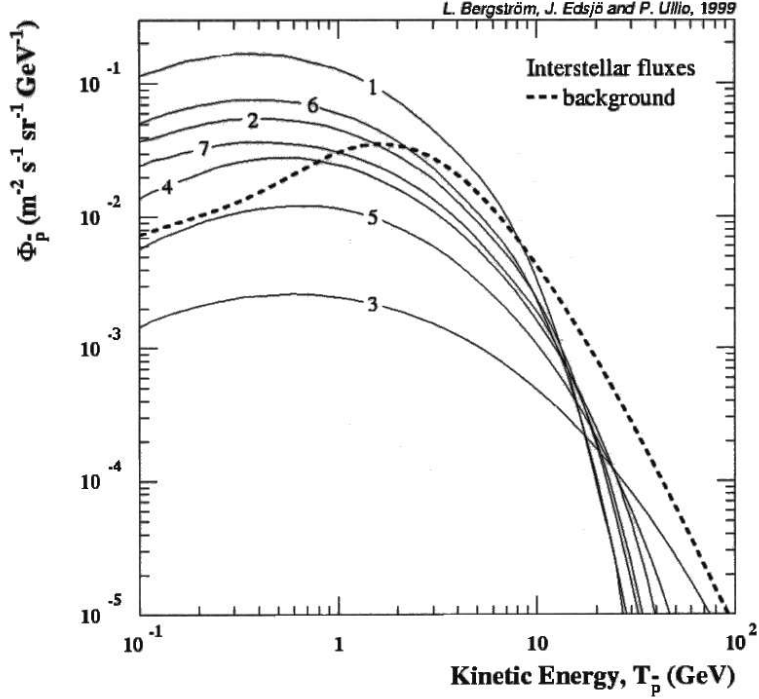} % postscript image file name
\caption{A range of minimal (R-parity-conserving) SUSY predictions of Galactic $\pbar$ spectrum,
with neutralino mass $m_{\tilde{\chi}^0}$ = 45--700 GeV and standard secondaries = ``background''.
Adapted from work of Bergstr\" om, Edsj\" o, and Ullio.~\protect\cite{Ullio}
\label{fig:ullio}}
\end{figure}

Although the secondary $\pbar$ spectrum must be there, its presence does not
rule out non-standard $\pbar$ sources, so-called ``exotic primaries''.  These
would add to the predicted secondary flux in total number.  More crucially,
they can also change the {\it shape} of the $\pbar$ spectrum, particularly
at low $K$, as well the fall-off for $K\gtrsim$ 3 GeV.  Cosmologically significant
amounts of antimatter are strongly disfavored.~\cite{CosmicAntimatter}  Instead,
the most logical sources for trace amounts of exotic primary $\pbar$ would be annihilating or decaying dark
matter remnants in the halo of our Galaxy.  
Popular models feature annihilating supersymmetric (SUSY) 
dark matter~\cite{CDMgeneral,LSP,CDMneutralino,CDMdetect}
(WIMPs, assumed to be the LSP = lightest SUSY particle, usually neutralinos $\tilde{\chi}^0$) or decaying primordial 
black holes.~\cite{Hawking,Stephens}  The predictions depend on model 
details,~\cite{Chard,Ullio,PBHs} but both have roughly flat $\pbar$ production spectra as 
$K\rightarrow$ 0 and a non-standard fall-off with $K$ at high energies.  
Such signals can only be seen if the exotic primaries compete in number with standard 
secondaries.  A general range of exotic SUSY $\pbar$ production (Figure~\ref{fig:ullio})
exhibits the dramatic modification of the low-$K$ $\pbar$ spectrum possible 
in SUSY CDM models of the Galactic halo for smaller neutralino mass.

\begin{figure}
%\figurebox{20pc}{15pc}{} % to have a box alone
\epsfxsize=20pc % will enlarge or reduce the postscript figures based on the xsize
\epsfbox{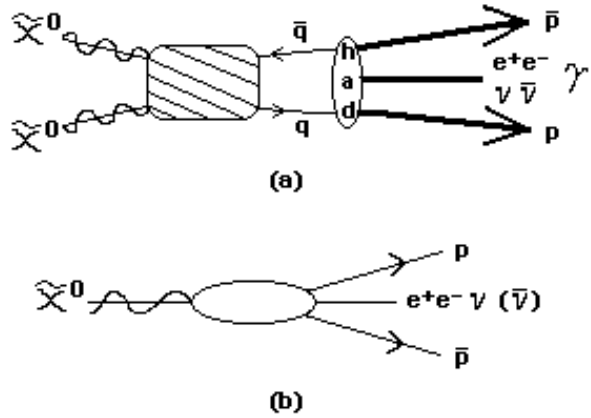} % postscript image file name
\caption{Supersymmetric CDM LSPs as a source of CR $\pbar$'s.  (a) CDM neutralinos 
$\tilde{\chi}^0$'s annihilate in the Galactic halo in minimal SUSY.  (b) In R-parity-violating
SUSY, neutralinos $\tilde{\chi}^0$'s can also decay by $\Delta L\neq$ 0 
channels.\label{fig:susyCDM}}
\end{figure}

\begin{figure}
%\figurebox{20pc}{15pc}{} % to have a box alone
\epsfxsize=25pc % will enlarge or reduce the postscript figures based on the xsize
\epsfbox{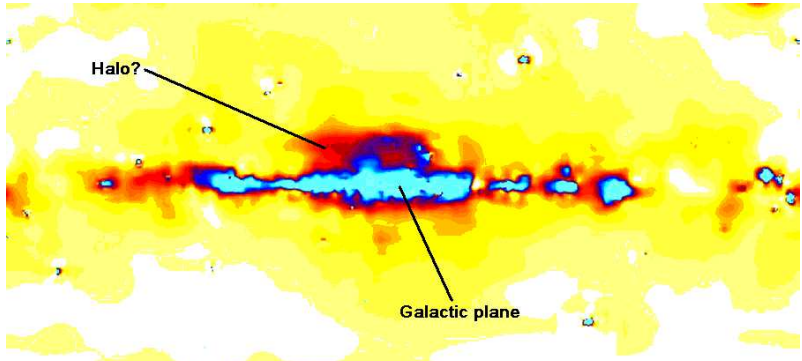} % postscript image file name
\caption{Intensity map of Galactic $\gamma$ rays: most intense is central, horizontal Galactic
plane, with possible fainter Galactic halo emission.  EGRET/Compton Gamma Ray 
Observatory.~\protect\cite{EGRET}\label{fig:egret}}
\end{figure}

SUSY halo dark matter $\pbar$'s (from $\tilde{\chi}^0\tilde{\chi}^0\rightarrow q\overline{q}$, 
Figure~\ref{fig:susyCDM}a) requires 
sufficient abundance and a large enough annihilation cross section to be seen,
in turn implying WIMP masses $\lesssim$ few 100 GeV and $\sigma ({\rm ann})
v({\rm WIMP})\sim$ 0.1 pb.  (Production of heavier WIMPs is suppressed in the 
Big Bang with with increasing mass.)  The hadronic shower evolves finally into $p$'s, 
$\pbar$'s, $e^\pm$, $\nu$'s, and $\gamma$'s.  WIMP annihilation is natural in minimal SUSY 
models with conserved $R$-parity.  An extension of minimal SUSY allows $R$-parity
violation, in turn allowing the LSPs to decay to ordinary matter, 
violating lepton and/or baryon number (Figure~\ref{fig:susyCDM}b).  
This mechanism could open another
source of CR $\pbar$'s.  In a semi-realistic scenario,~\cite{Dreiner}
lepton number violation is dominant, leading in the end to excess (anti)neutrinos.
An exciting possible signal of annihilating or decaying CDM in the Galactic halo is
suggested by the Galactic gamma ray maps of the orbiting Compton Gamma Ray Observatory's 
EGRET telescope (Figure~\ref{fig:egret}).

Primordial black holes (PBHs) are postulated to have been produced very early in the hot 
Big Bang, in the quantum gravity era.~\cite{Hawking}  They evaporate in turn by the
Hawking process, as their temperatures rise, and can produce significant 
$p$'s and $\pbar$'s at a late time when $T_{\rm BH}\gtrsim$ $\Lambda_{\rm QCD}$.  The relic 
PBH density and $\pbar$ production rate have been estimated.~\cite{PBHs}

\subsection{Intrinsic Properties of Antimatter: CPT Symmetry}\label{sec:cpt}

Cosmic ray $\pbar$'s also give us a window on the intrinsic properties of 
antimatter.  These properties should be the same or charge-conjugated from the 
corresponding matter by the CPT (charge-conjugation, parity- and time-reversal)
symmetry of local relativistic quantum field
theory (LRQFT).  Some $\pbar$ properties have been checked in the laboratory 
directly.  These include the mass, charge, magnetic moment, and the neutrality
of hydrogen and antihydrogen.~\cite{PDBppbar}

More difficult to limit is the decay lifetime of $\pbar$'s.  Not enough
antimatter can be gathered into a detector for long enough to produce lifetime
limits on antimatter competitive with the limits for matter.
Astrophysical processes partially ameliorate this difficulty.  The
Galactic storage time for $\pbar$'s $\sim$ 10 Myr, and intrinsic $\pbar$ decay
would modify the Galactic residence time $\tau_{\rm eff}$~{}.  If the decay lifetime
is short enough (taking Lorentz dilation into account), the $\pbar$ spectrum
is significantly distorted.  The shape and normalization of the $\pbar$
spectrum then place a lower limit on $\tau_{\pbar}$.~\cite{GeerKenn}

Laboratory limits have been obtained for the $\pbar$ lifetime.  Earlier limits
include the LEAR Collaboration at the CERN $\pbar$ storage ring $(\tau_{\pbar}
>$ 0.08 yr) and the antihydrogen Penning trap of Gabrielse {\it et al.}
$(\tau_{\pbar}>$ 0.28 yr).~\cite{PDBppbar}  The best current laboratory limit is that of the
APEX Collaboration at the Fermilab $\pbar$ storage ring $(\tau_{\pbar}>$ 50
kyr for $\pbar\rightarrow\mu^- X$ and 300 kyr for $e^-\gamma$).~\cite{APEX}  A proposed
APEX II experiment would be able to reach $\pbar$ lifetime limits of 1--10
Myr, comparable to the cosmic ray limit.\footnote{All lifetime limits quoted
here are at 90\% C.L.}

Since the CPT symmetry holds in LRQFT under the assumptions of Poincar\' e invariance, locality, 
microcausality, and vacuum uniqueness, modification of basic physics would be 
necessary to break it.~\cite{CPTthm,LorentzCPT}
Within QFT, an extensive formalism and phenomenology of Lorentz and CPT violation has 
been developed by Kosteleck\' y and collaborators.~\cite{KostelCPT}  String theory at first 
glance might seem to provide a natural way to violate locality, but perturbative string 
dynamics has been shown to preserve CPT in the field theory target space after 
compactification.~\cite{KostelStrings}  {\em Non-perturbative} string effects associated with
compactification may evade this result.~\cite{Polchin}
Extended quantum mechanics, with non-unitary 
time evolution, violates CPT in general, by violating locality and/or Poincar\' e symmetry.  
Controversial proposals of non-unitary evolution have been put forward as natural consequences 
of quantum gravity and information loss in the presence of spacetime horizons.~\cite{EHNScpt}  
Non-unitary effects have been powerfully limited in the very well-measured 
$K^0$--$\overline{K}^0$ system~\cite{CPLEAR} (to a few parts in $10^{16}$), but not well at all 
in other systems, particularly baryons.~\cite{KennCPT}

The most plausible source of CPT violation lies beyond the Planck scale, based on strings or
some other quantum theory of gravity, because of the necessary generalization beyond global
Poincar\' e symmetry.  Typically such effects are thought of as suppressed by the large Planck 
mass $M_{\rm Pl}\sim 10^{19}$ GeV.  But if gravity is fundamentally associated with ``large''
extra dimensions acting at mass scales as low as 1 TeV,~\cite{LargeExtraDim} the CPT-violating
mechanisms may not be that suppressed at accessible energies.

\section{Measurements of Cosmic Ray Antiprotons}

Detection of CR antiprotons has gone through three distinct phases, following the
proposal of Gaisser and Levy to search for $\pbar$ secondaries.~\cite{GaisLevy}
All but recent space-based experiments have been mounted on high-altitude
balloons.  The measurements are
conventionally quoted as the $\pbar /p$ ratio of fluxes, convenient because
a number of theoretical and experimental uncertainties cancel: the overall
IS primary $p$ flux normalization uncertainty, the overall detector flux
normalization uncertainty, and (at $K\gtrsim$ 500 MeV) diffusive modulation
of both fluxes (see below).

The first two Western experiments (those of Golden {\it et al.} and Buffington
{\it et al.}) detected $\pbar$ signals at a level higher than the standard 
secondary prediction.~\cite{Golden,Buffington}  These early experiments detected $\pbar$'s
by energy calorimetry (the deceleration and annihilation of the $\pbar$'s in 
the balloon), but lacked definite identification by a magnetic spectrometer.  
Of particular concern is the background of kaons in the detector, as $m_K\lesssim m_p$.

Stimulated by the possibility of an excess of CR $\pbar$'s, a number of groups
completed measurements in the 1970s and 1980s with better particle 
identification.  The PBAR and LEAP groups established upper limits on the
CR $\pbar$ flux contradicting the first-generation experiments.~\cite{PBAR,LEAP}  Roughly
contemporaneous, the Soviet group of Bogomolov {\it et al.} reported three flux
measurements (from the periods 1972-77, 1984-85, and 1986-88) 
consistent with standard secondary predictions.~\cite{Bogo1,Bogo2,Bogo3}

\subsection{Abundance \& Spectrum of Antiprotons}

The third generation of experiments came in the 1990s and included markedly better particle 
detection by magnetic spectrometer, of quality comparable to accelerator experiments.  
From 1991 to 1997, the MASS (1991), IMAX (1992), CAPRICE (1994), and BESS (1993, 1995, 1997) 
collaborations have made clean measurements of the CR $\pbar$ flux with low 
backgrounds.~\cite{MASS91a,IMAX,BESS93,CAPRICE94,BESS95,BESS97}

\begin{table}
\begin{center}
\caption{Summary of cosmic ray $\pbar$ measurements not contradicted by later experiments.
Adapted from Geer and Kennedy.~\protect\cite{GeerKenn}\label{tab:Summary}}
\eightrm
\begin{tabular}{|lc|c|c|c|c|c|c|c|}
\hline
Experiment& &Field&Flight&KE Range&Cand-&Back-&Observed 
            &Predict-\\
            & &Pol.$^{\rm a}$&Date&(GeV)&idates&ground&$\pbar /p$ Ratio&
            tion$^{\rm b}$\\
\hline
Golden et al. 1979$^\dag$&&$+$&June 1979&5.6 -- 12.5&46&18.3&
$(5.2\pm 1.5)\times 10^{-4}$& -- \\ 
\hline
%Buffington$^\dag$&\cite{buff}&${\rm n/a}$&June 1980&0.13 -- 0.32&17&3.0&
%$(2.2\pm 0.6)\times 10^{-4}$& -- \\ 
%\hline
Bogomolov et al. 1979$^\dag$&&$+$&1972-1977&2.0 -- 5.0&2&--&
$(6\pm 4)\times 10^{-4}$& -- \\
Bogomolov et al. 1987$^\ddag$&&$-$&1984-1985&0.2 -- 2.0&1&--&
$(6^{+14}_{-5})\times 10^{-5}$& -- \\
Bogomolov et al. 1990$^\ddag$&&$-$&1986-1988&2.0 -- 5.0&3&--&
$(2.4^{+2.4}_{-1.3})\times 10^{-4}$& -- \\ \hline
MASS91~\protect\cite{MASS91a}&&$+$&Sep. 1991&3.70--19.08&11&3.3&
$(1.24^{+0.68}_{-0.51})\times 10^{-4}$&$1.3\times 10^{-4}$\\
\hline
IMAX$^{\ddag}$~\protect\cite{IMAX}&&$+$&July 1992&0.25 -- 1.0&3&0.3&
$(3.14^{+3.4}_{-1.9})\times 10^{-5}$&$1.5\times 10^{-5}$\\
IMAX~\protect\cite{IMAX}&&$+$&July 1992&1.0 -- 2.6 &8&1.9&
$(5.36^{+3.5}_{-2.4})\times 10^{-5}$&$6.5\times 10^{-5}$\\
IMAX~\protect\cite{IMAX}&&$+$&July 1992&2.6 -- 3.2 &5&1.2&
$(1.94^{+1.8}_{-1.1})\times 10^{-4}$&$1.1\times 10^{-4}$\\
\hline
BESS93$^{\ddag}$~\protect\cite{BESS93}&&$+$&July 1993&0.20 -- 0.60&7&$
\sim 1.4$&$(5.2^{+4.4}_{-2.8})\times 10^{-6}$&$8.9\times 10^{-6}$\\
\hline
CAPRICE~\protect\cite{CAPRICE94}&&$+$&Aug. 1994&0.6 -- 2.0 &4&1.5&
$(2.5^{+3.2}_{-1.9})\times 10^{-5}$&$3.5\times 10^{-5}$\\
CAPRICE~\protect\cite{CAPRICE94}&&$+$&Aug. 1994&2.0 -- 3.2 &5&1.3&
$(1.9^{+1.6}_{-1.0})\times 10^{-4}$&$1.1\times 10^{-4}$\\
\hline
BESS95$^{\ddag}$$^\ast$~\protect\cite{BESS95}&&$+$&July 1995&0.175 -- 0.3&3&0.17&
$(7.8^{+8.3}_{-4.8})\times 10^{-6}$&$-$\\
BESS95$^{\ddag}$$^\ast$~\protect\cite{BESS95}&&$+$&July 1995&0.3 -- 0.5&7&0.78&
$(7.4^{+4.7}_{-3.3})\times 10^{-6}$&$1.1\times 10^{-5}$\\
BESS95$^{\ast}$~\protect\cite{BESS95}&&$+$&July 1995&0.5 -- 0.7&7&1.4&
$(7.7^{+5.3}_{-3.7})\times 10^{-6}$&$5.5\times 10^{-6}$\\
BESS95$^{\ast}$~\protect\cite{BESS95}&&$+$&July 1995&0.7 -- 1.0&11&2.8&
$(1.01^{+5.7}_{-4.3})\times 10^{-5}$&$1.3\times 10^{-5}$\\
BESS95$^{\ast}$~\protect\cite{BESS95}&&$+$&July 1995&1.0 -- 1.4&15&3.5&
$(1.99^{+0.91}_{-0.73})\times 10^{-5}$&$3.1\times 10^{-5}$\\
\hline
\end{tabular}
\end{center}
\vspace{0.25cm}
\indent{${}^{\rm a}$~Northern hemisphere heliomagnetic polarity:
+ = outward field.}\\
%\indent{${}^{\rm b}$~Epoch correction factor: see text.}\\
\indent{${}^{\rm b}$~ILDM prediction with $V_W({\rm eq/polar})$ = 400/750
km sec$^{-1}$, $B_\odot$ = 4.5 nT.}\\
\indent{${}^\dag$~Not shown in Figure~\ref{fig:unmodRatio} or used in analysis.
${}^\ddag$~Not used in analysis.  
${}^\ast$~Statistical and systematic uncertainties on ratio 
added in quadrature.}
\label{balloon_tab}
\end{table}

\begin{figure}
%\figurebox{20pc}{15pc}{} % to have a box alone
\epsfxsize=25pc % will enlarge or reduce the postscript figures based on the xsize
\epsfbox{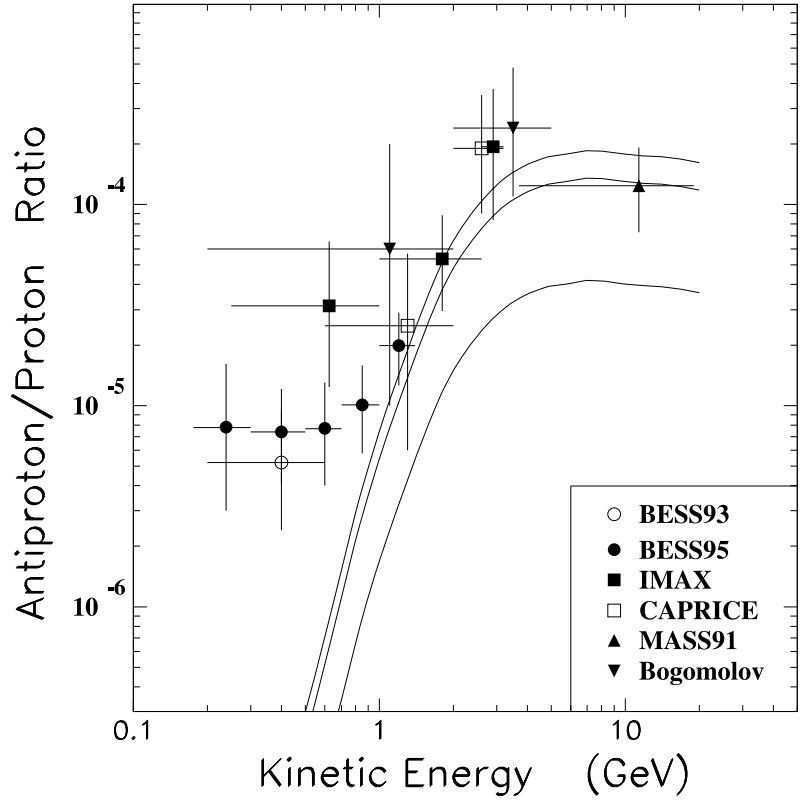} % postscript image file name
\caption{Measured $\pbar /p$ spectral flux ratios, compared with {\em unmodulated} IS CR
$\pbar$ flux predicted by ILDM (see text).~\protect\cite{GeerKenn}\label{fig:unmodRatio}}
\end{figure}

\begin{figure}
%\figurebox{20pc}{15pc}{} % to have a box alone
\epsfxsize=25pc % will enlarge or reduce the postscript figures based on the xsize
\epsfbox{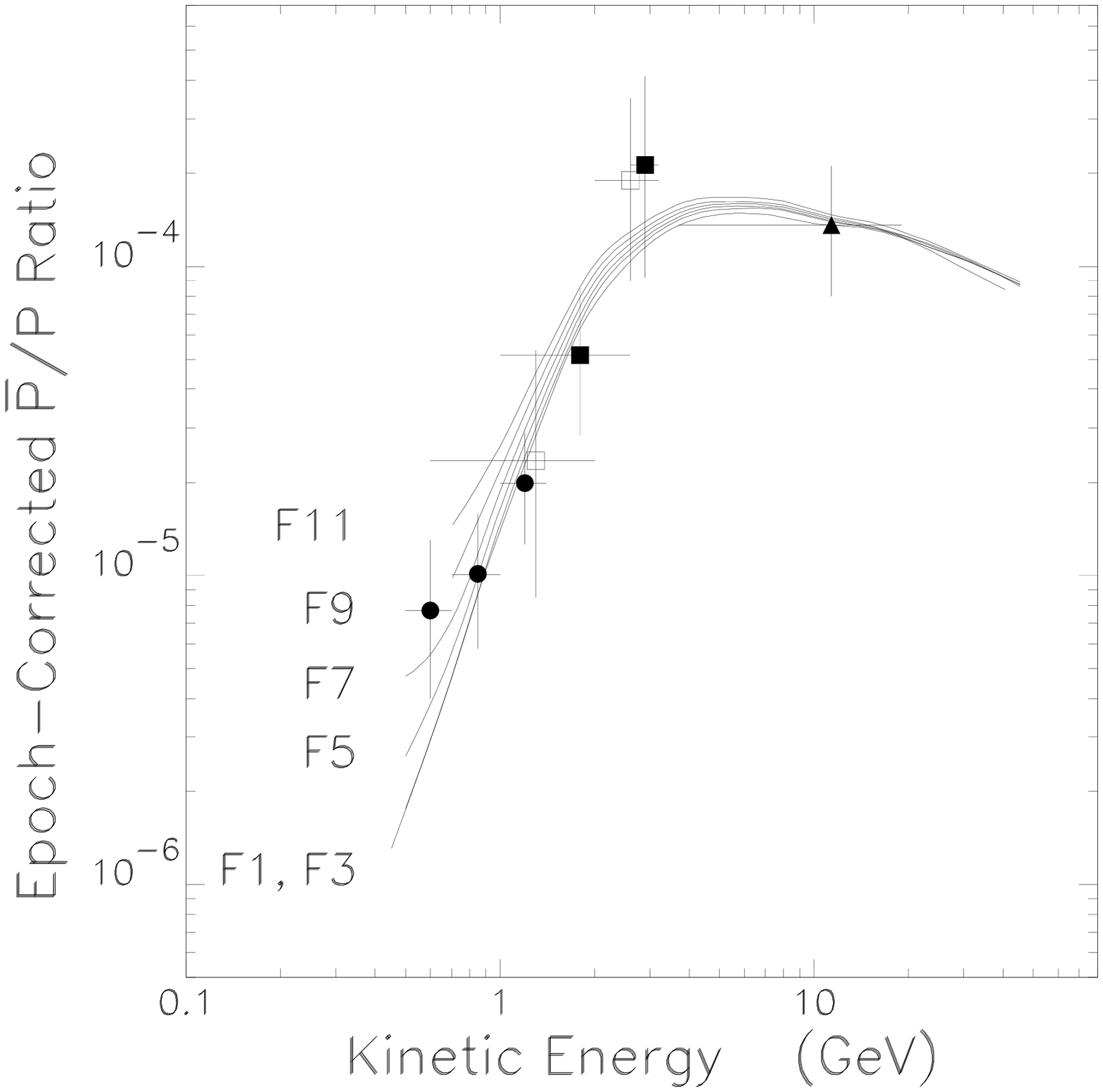} % postscript image file name
\caption{Measured $\pbar /p$ spectral flux ratios corrected with heliospheric modulation
to common epoch (July 1995), compared with modulated prediction of IS CR $\pbar$ flux in
ILDM (see text).~\protect\cite{GeerKenn}\label{fig:modRatio}}
\end{figure}

The analysis presented here in based on all refereed and published measurements not 
contradicted by later measurements with better detectors (Table~\ref{tab:Summary}).~\cite{GeerKenn}
Figure~\ref{fig:unmodRatio} shows the selected measurements compared with the ILDM prediction for IS
fluxes.  The disagreement evident in the figure is explicable by heliospheric
modulation.  Figure~\ref{fig:modRatio} compares the modulated ILDM predictions with the
measured fluxes.  This figure makes the comparison by renormalizing the
measured fluxes to a single epoch (July 1995, chosen as roughly the most
recent heliomagnetic minimum) and using the prediction for that epoch.  Our
analysis did not use measurements with $K <$ 500 MeV because of the large and 
difficult-to-calculate diffusion modulation in that energy range.

\subsection{Implications~\protect\cite{GeerKenn,KennCPT}}

The most basic result implied by Table~\ref{tab:Summary} and Figure~\ref{fig:modRatio}
is the standard $\pbar$ secondary flux alone, from a realistic ILDM, can account for the 
observed flux in the relevant energy range, within uncertainties.  If variant Galactic
transport mechanisms (such as reacceleration or shrouded sources~\cite{Gaisser}) or exotic
$\pbar$ sources are at work in this $K$ range, their effects are too small to see at this
time.  (A hint of reacceleration may be visible in the range $K\simeq$ 2--5 GeV by
distortion of the spectrum evident in Figure~\ref{fig:modRatio}, but the effect is not
significant within uncertainties.)
A second, less obvious, result is a limit on the intrinsic decay lifetime of the
antiproton: $\tau_{\pbar} >$ 0.8 Myr, the best limit currently feasible.
While the exclusion of the $K <$ 500 MeV spectrum does {\em not}
significantly affect the $\tau_{\pbar}$ limit, it does limit conclusions about the absence
of exotic $\pbar$ sources, as these would have their largest effect relative to the standard
secondaries precisely at such low $K$.

\begin{table}
\caption{$CPT$-- and $B$--violating scale limits associated with $p$ lifetime $\taup$ 
= $10^{32}$ yr and $\pbar$ lifetime $\tpbar$ = $10^7$ yr (see text).~\protect\cite{KennCPT}
\label{tab:CPTscales}}
\vspace{0.2cm}
\begin{center}
\footnotesize
\begin{tabular}{|c|c|c|c|}
\hline
$n$ & $M_X$ (GeV) & $k$ & $M_Y$ (GeV) \\
\hline
   &                   &   &                  \\
 5 & $2\times 10^{19}$ & 1 & $2\times 10^{63}$\\
 6 & $4\times 10^9$ &    2 & $5\times 10^{31}$\\ 
 7 & $3\times 10^6$ &    3 & $1\times 10^{21}$\\
 8 & $6\times 10^4$ &    4 & $7\times 10^{15}$\\
 9 & $7\times 10^3$ &    5 & $5\times 10^{12}$\\
10 & $2\times 10^3$ &    6 & $4\times 10^{10}$\\
\hline
\end{tabular}
\end{center}
\end{table}

A short $\pbar$ lifetime $\tau_{\pbar}\lesssim$ 10 Myr (Galactic CR storage time) would of
course indicate CPT violation.  The two pictures of CPT violation introduced in 
subsection~\ref{sec:cpt} are: modification of LRQFT within ordinary quantum mechanics, and
non-standard quantum mechanics (NSQM) with non-unitary time evolution.  If only one new
mass scale is relevant to the CPT violation,
lower limits can be placed on such scales.  In Table~\ref{tab:CPTscales}, the limiting 
CPT-violating scales associated with modified QFT $(M_X)$ and NSQM $(M_Y)$ are
shown, assuming $\tau_{\pbar}$ = 10 Myr.  The $\pbar$ lifetime is assumed related to each
scale by simple mass dimensions.  For modified QFT, $\Gamma_{\pbar}$ = $m_p(m_p/M_X)^n$; while
for NSQM, $\Gamma_{\pbar}$ = $(m_p/2)(m_p/M_Y)^k$.  It is interesting to note that the {\em largest}
$M_X$ lower bound is ${\cal O}(M_{\rm Pl})$, while the scales of order the ``intermediate''
scale $(10^8$--$10^{12}$ GeV) are possible, as well as scales $\sim$ TeV.  The last scale may
not be unreasonably low in the context of ``large'' extra dimensional 
gravity.~\cite{LargeExtraDim}

\section{Future Developments and Prospects}

Uncertainties intrinsic to cosmic ray analysis will probably limit deduction of antimatter 
properties to about the level already achieved.  But the search for exotic sources of
primary $\pbar$'s is still open, especially at low energy.

\subsection{More and Better Measurements}

Future measurement of the medium energy range $(K$ = 0.5--10 GeV)
will define that part of the spectrum better, but it the spectral shape at the two
extremes that is critical for exotic $\pbar$ searches.

A number of experiments have already taken recent data not yet published.
These include the CAPRICE (1998) and HEAT (1999) balloons, as well as the prototype AMS (1998)
and PAMELA (1995 and 1997) systems tested on Space Shuttle STS-91 and the Mir space station,
respectively.~\cite{SoonPubExps,UpcomingExps}  These experiments can and have searched for
positrons and $A >$ 1 antinuclei as well.   The HEAT-$\pbar$99 data are especially of interest
because of their large energy range ($K$ = 4--50 GeV).

The PAMELA instrument, after being tested in prototype
on the Mir space station, is scheduled to fly on an unmanned satellite (the Russian-Italian
Resurs-Arktika 4) for three years, starting in 2002.~\cite{UpcomingExps}  It can detect $e^+$, $\pbar$, and
$\overline{\rm He}$ at a relative sensitivity of better than one part in $10^7$ over a range
$K$ = 0.1--150 GeV.  The full AMS instrument is scheduled for the International Space Station
Alpha starting in 2005, also for three years, with an antiparticle/antinucleus sensitivity of one
part in $10^6$ for $E >$ 5 GeV.

The MASS91 collaboration have also reanalyzed their data and released a new version divided into
three energy bins, instead of one.~\cite{MASS91b}  These three experiments (MASS91, HEAT,
and PAMELA) will decisively address the paucity of data at the highest energies and define
the spectrum in that range.

\subsection{Production \& Propagation: Importance of the Low-Energy 
Spectrum}\label{sec:FutureTheo}

The low-energy range is already being mapped out by the BESS experiment, in particular in
the 1995 and 1997 data sets.  Repeated, reliable measurement of the low-energy spectrum is 
the most important task in the contemporary period of
$\pbar$ measurements, followed closely by reliable measurement of the high-energy fall-off.
The presence of exotic primary $\pbar$'s in this range should be detectable with the current 
or next generation of experiments.

The main obstacles to conclusive limits on a non-standard $\pbar$ flux at low energy are now
theoretical.  There are two crucial effects needing clarification for such a signal to be
found or ruled out.  The first is the ``subthreshold'' $\pbar$ production on IS He-4 target 
nuclei.  The status of previous estimates of this effect~\cite{WebbPot,GaisSch} 
has been changed in the last decade
by laboratory measurements of the $\pbar$ production on heavy target nuclei,~\cite{Sibir}
providing evidence for a scaling relation between the $A >$ 1 and $A$ = 1 cases.
Recent calculations~\cite{Ullio} have begun to take account of these data, and further work
is under way to develop a simple nuclear model.~\cite{KennMillWP}

The second is providing a complete heliospheric modulation calculation that includes diffusion,
as well as the wind and magnetic drift.  The present gap in the literature is defined
on one side by thorough modulation calculations applied to low-energy CRs $(K <$ 100
MeV)~\cite{transportNew}
and on the other by accurate modulation done at higher energies $(K >$ 500 MeV) without
diffusion.~\cite{GeerKenn}  Approximate calculations without magnetic drift are 
available,~\cite{transportOld} but the charge-dependent magnetic drift is essential to 
predicting the $\pbar /p$ ratio correctly.  A full calculation covering $K\lesssim$ 100 MeV
to 500 MeV is essential to proper interpretation of the BESS data.~\cite{GeerKennWP}

Cosmic ray antimatter measurements are undergoing exciting developments 
that will define much of our future
understanding of the composition of our Galaxy and of basic symmetries of Nature.  Perhaps
within 10 years, precise cosmic ray measurements will be a mature subject, along
with the ripening of other types of particle astrophysics.

\section*{Acknowledgments}
The author thanks the PASCOS 99 meeting organizers at UC Davis
for the opportunity to present these results, based on work done in collaboration
with Stephen Geer (Fermilab).  This work was supported by the 
U.S. DOE under grants DE-FG02-97ER41029 (Univ. Florida) 
and DE-AC02-76CH03000 (Fermilab), NASA under grant NAG5-2788 (Fermilab),
and the Institute for Fundamental Theory at the Univ. Florida, and was
greatly enhanced by conversations with J.~.R.~Jokipii (Lunar \& Planetary Laboratory, 
Univ. Arizona), E.~J.~Smith (Jet Propulsion Laboratory),
V.~A.~Kosteleck\' y (Indiana Univ.), and M.~Kamionkowski (CalTech).
The author is also grateful to the NASA/Fermilab Theoretical Astrophysics group and the
Telluride Summer Research Center for their hospitality.

\end{document}